\documentclass[11pt, a4paper]{article}
\usepackage{jheppub}
\usepackage{commath}
\usepackage{braket}
\usepackage{dsfont}
\usepackage{amsmath}
\usepackage{amsfonts}
\usepackage{graphicx}

\usepackage{bbold}

\usepackage{parskip}

\newcommand*{\bfn}{{ \bf n }}

\newcommand*{\bfei}{{ \bf e_i }}

\begin{document}

\title{Continuum limit of gauged tensor network states}
\begin{abstract}{
 	It is well known that all physically relevant states of gauge theories lie in the sectors of the Hilbert space which satisfy the Gauss law. On the lattice, the manifeslty gauge invariant subspace is known to be exactly spanned by gauged tensor networks. In this work, we demonstrate that the continuum limit of certain types of gauged tensor networks is well defined and leads to a novel class of states. 
    
    We discuss several methods to compute generating functionals that may be helpful for the non-perturbative study of gauge theories directly in the continuum. 
}\end{abstract}

\author[a]{Gertian Roose}
\author[a]{Erez Zohar}
\emailAdd{Gertian.Roose@gmail.com}
\affiliation[a]{School of Physics and Astronomy, Tel Aviv University, Tel Aviv 6997801, Israel}

\date{\today}

\maketitle

\section{Introduction}
Gauge theories form one of the backbones of modern physics. Most notably, in the standard model they are responsible for the fundamental interactions between matter \cite{Peskin_an_introduction_to_quantum_field_theory} and more recently, it has been demonstrated that they play a key role in describing higher form symmetries \cite{Gomes_an_introduction_to_higher_form_symmetries,Seiberg_Generalized_global_symmetries} as well as dualities on the lattice \cite{Bram_From_gauging_to_duality_in_one-dimensional_quantum_lattice_models,Ashkenazi_Duality_as_a_feasible_physical_transformation_for_quantum_simulation} and in the continuum \cite{Roose_T-duality_and_bosonization_as_examples_of_continuum_gauging_and_disentangling,Busher_Path_integral_derivation_of_quantum_duality_in_nonlinear_sigma_models}. Despite their huge importance, gauge theories are notoriously hard to study due to their rich phase diagram that contains many phases that cannot be understood perturbatively \cite{reddy_novel_phases_at_high_density_and_their_roles_in_the_structure_and_evolution_of_neutron_stars, RAJAGOPAL_2001}. 

Indeed, the gauge invariant objects that constitute relevant states are non-local Wilson loops and lines that are not well described within perturbation theory when the coupling is large \cite{Gross_Asymptotically_free_gauge_theories_I}. To better understand strongly interacting gauge theories it is therefore desirable to have a framework that contains only these gauge invariant objects from the start, on the lattice it has been demonstrated that gauged projected entangled pair states (gPEPS) \cite{Zohar_Combining_tensor_networks_with_Monte_Carlo_methods_for_lattice_gauge_theories,Zohar_2016, Haegeman_2015} provide such a framework. 

For pure matter tensor network states (TNS), the continuum limit of tensor network states is well known to be continuous matrix product states (CMPS) \cite{Verstraete_2010, Haegeman_2013} that have proven to be a valuable numerical ansatz \cite{PhysRevLett.128.020501, Haegeman_2010, Ganahl_2017, PhysRevB.95.045145, PhysRevB.90.235142, PhysRevA.92.043629, Tiwana2025, tilloy2022studyquantumsinhgordonmodel}. In higher dimensions this continuum limit is given by continuous projected entangled pair states (CPEPS) \cite{Shachar:2021vbu, Tilloy_2019, brockt2012continuumlimittensornetwork} that have also been used as a numerical ansatz \cite{Karanikolaou_2021}. 

In this work, we extend this program of taking continuum limits of tensor network states to the case of gauged projected entangled pair states. This leads to gauged continuous projected entangled pair states (gCPEPS). Naturally, when the gauge theory coupling constant $g=0$, the proposed state reduces to the conventional CPEPS introduced in \cite{Tilloy_2019} and for such states, it is known that the generating functionals in $D \leq 3$ are convergent if the generating Lagrangian of the state is Gaussian. In the same work, it was argued that other cases can be dealt with through usual QFT tricks. For example, for non Gaussian generating functionals in $D \leq 3$, one can expand perturbatively around the Gaussian scenario and renormalize the two point function using a UV-cutoff or with the addition of counter-terms in the generating Lagrangian. In our case, it is natural to express this perturbative renormalisation in the coupling $g$. Alternatively when $D > 3$ or when perturbation theory is not sufficient one can apply a tensor network inspired contraction scheme to iteratively reduce the dimension of the problem down to lower dimensions. Note that the existence of this continuum limit is particularly relevant in light of the recent proofs that gauged PEPS and MPS are the most general gauge invariant states on the lattice \cite{Kull_Classification_of_matrix_product_states_with_a_local_gauge_symmetry,Blanik_Internal_structure_gauge_invariant_PEPS}.

The outline of the paper is as follows. In section 2 we start by defining gauged continuous tensor networks in a way that is explicitly Lorentz invariant. In subsection 2.1 we explicitly demonstrate gauge invariance of these states and in subsection 2.2 we introduce an alternative representation that is more suitable for computations. In subsection 2.3 we discuss a natural simplification in 1 dimension. In section 3 we introduce generating functionals and discuss how to compute them for our states. More specifically, in subsection 3.1 we focus on states that are perturbatively close (in the gauge theory coupling $g$) to Gaussian continuous tensor networks and in subsection 3.2 we discuss a numerical method that is valid more generally.  Throughout these sections we discuss renormalization where relevant. Finally, in section 4 we demonstrate that gauged continuous PEPS are really the continuum limit of sequences of discrete gauged tensor networks. To do this we start in subsection 4.1 with a brief review of the relevant pure matter tensor networks on the lattice. In subsection 4.2 we demonstrate how to gauge those and finally in subsection 4.3 we take the continuum limit. 

\section{Gauged continuous tensor network states}
Given a gauge group G with irreducible representation (irrep) labels $j$ and color labels $\alpha$, we wish to construct manifestly gauge invariant states on a spatial manifold $\mathcal{M}$ with boundary $\partial \mathcal{M}$. Here we focus on gauge theories with matter fields that are complex scalars $\hat\phi_\alpha(x)$ in the fundamental representation $j_f$ of G and gauge fields $\hat A_\mu^a(x)$ that are valued in the lie-algebra of G. Apart from these, our construction depends on a set of auxiliary complex scalar fields $\hat \chi^{j}_{\alpha}(x)$ in the j-th representation of G. Naturally, each of these fields has a canonically conjugate momentum, in particular for the gauge fields we denote $\hat \Pi_{A^a_\mu }(x) = \hat E^a_\mu(x)$ so that $ [\hat A^a_\mu(x), \hat E^b_\nu(y)] = i \delta(x-y)\delta^{a b}\delta_{\mu \nu} $. Additionally, we define creation and annihilation operators for the physical fields i.e.:
\begin{align}
	\hat{a}_\alpha(x) &= \frac{1}{\sqrt{2}}\del{ \sqrt{\Lambda} \hat \phi_\alpha(x) + i \frac{1}{\sqrt{\Lambda}}\hat \Pi^\dagger_\alpha(x) } \\
	\hat{b}^{\dagger}_\alpha(x) &= \frac{1}{\sqrt{2}}\del{\sqrt{\Lambda}\hat \phi_\alpha(x) - i \frac{1}{\sqrt{\Lambda}} \hat \Pi^\dagger_\alpha(x)  }  \label{eq:phys_particles} 
\end{align}
where $\hat\Pi_\alpha(x)$ is the conjugate momentum of $\hat\phi_\alpha(x)$ and $\Lambda$ is some arbitrary dimensionful parameter required for consistency. Finally, we define the Fock vacuum $\ket{0}$ of the creation operators and the electric singlet state $\ket{s_E}$ that satisfies $\hat E^a_\mu(x) \ket{s_E} = 0$. 

Using all the above definitions we now introduce the variational parameters of the state in the form of two bulk functionals $V[\chi, \partial_\mu \chi, \bar \chi, \partial_\mu \bar \chi]$, $J[\chi, \partial_\mu \chi, \bar \chi, \partial_\mu \bar \chi, \hat{a}^{\dagger \, j}, \hat{b}^{\dagger \, j} ]$ and one boundary functional $B[\chi_{\partial \mathcal{M}}, \partial_\mu \chi_{\partial \mathcal{M}}, \bar \chi_{\partial \mathcal{M}}, \partial_\mu \bar \chi_{\partial \mathcal{M}}]$ that depend on creation operators of the physical field and the virtual fields their derivatives and the complex conjugates thereof. Using those, the gauge invariant CPEPS is defined as:
\begin{align}
    \ket{\psi_{B, V, J}} = \int \mathcal{D}^2 \chi \, B[\chi_{\partial M}] e^{ -\int_\mathcal{M} d^D x \hat{\mathcal{L}}[\hat A^a_\mu ,\chi, \hat a^\dagger, \hat b^\dagger]   } \ket{ 0 } \ket{s_E} \label{eq:gauged_CPEPS}
\end{align}
where $\mathcal{D}^2 \chi = \mathcal{D} \chi \mathcal{D} \bar \chi$ and similar for the auxiliary fields. The generating Lagrangian is defined as: 
\begin{align}
    \hat{\mathcal{L}}[\hat A^a_\mu ,\chi, \hat a^\dagger, \hat b^\dagger] = \del{\hat D_\mu \chi^{j }_{\alpha }(x)}^\dagger  \hat D^\mu \chi^{j}_{\alpha }(x) + V[\chi] + J[\chi, \hat a^\dagger, \hat b^\dagger]    
\end{align}
and contains the covariant derivative $\hat D_\mu$ that is defined as:
\begin{align}
    \hat D_\mu \chi^{j}_{\alpha}(x)= \partial_\mu \chi^{j}_{\alpha}(x)- i g \hat A_\mu^a(x) \ (T^{j\,a})_{\alpha  \beta} \ \chi^{j}_{\beta}(x) \label{eq:covariant_derivative}
\end{align} 
where $g$ is the dimensionful coupling constant of the relevant gauge theory and $T^{j \, a}$ are the generators of the group acting on the irrep labeled by $j$. 

As we will soon see and discuss in more detail, the above described state is gauge invariant if the generating Lagrangian $\hat{\mathcal{L}}[\hat A^a_\mu ,\chi, \hat a^\dagger, \hat b^\dagger]$ is that of an interacting gauge theory for gauge group G and the boundary functional $B[\chi_{\partial M}]$ is a singlet.  A natural choice for the boundary condition that satisfies this is given by $B[\chi_{\partial \mathcal{M}}] \propto \delta(\chi_{\partial \mathcal{M}} \bar \chi_{\partial \mathcal{M}} )$. Other, non-local, choices are possible and discussed in \cite{Tilloy_2019}. 

\subsection{Explicit verification of gauge invariance and Gauss law}
Here we wish to explicitly check that $\ket{\psi_{B, V, J}}$ satisfies the Gauss law constraints:
\begin{align}
    \hat G^a(x) \ket{\psi_{B,V,J}} = \del{ \hat Q^a(x) - \frac{1}{g} \hat D_\mu \hat{E}^a_\mu(x)  }\ket{\psi_{B,V,J}} = 0 \hspace{1cm} \forall\ a , \ x \label{eq:Gauss_law}
\end{align} 
for suitable choices of the parameters $B,V,J$. 

In the above, 
\begin{align}
    \hat{Q}^a(x) &= - i \, \hat \Pi_\alpha(x) \, (T^{j_f \, a})_{\alpha \beta} \, \hat \phi_\beta(x) + h.c. 
\end{align}
or
\begin{align}
    \hat{Q}^a(x) &=  a^\dagger_\alpha(x) T^{j_f \, a}_{\alpha \beta} a_\beta(x) -  b^\dagger_\alpha(x) (T^{j_f \, a}_{\alpha \beta})^T b_\beta(x) \label{eq:charge_operator}
\end{align}  
is the charge density of the scalar matter fields and
\begin{align}
    \hat D_\mu \hat{E}_\mu^a(x) = \partial_\mu \hat{E}_\mu^a(x) - ig \hat A^b_\mu(x) (i f^{abc}) \hat E^c_\mu(x)
\end{align} 
is the covariant derivative of the electric field. To relate the covariant derivative for the electric field to that of the virtual fields (c.f.r. \ref{eq:covariant_derivative}) one uses the fact that the electric field $E_\mu^a(x)$ is valued in the adjoint representation of G so that the relevant generators are 
\begin{align}
    (T_{\text{adjoint}}^b)_{ac} = i f^{abc}    
\end{align}
where $f^{abc}$ are the structure constants of the group. 

To prove \ref{eq:Gauss_law} we will instead prove the equivalent statement that $\ket{\psi_{B, V, J}}$ has a local symmetry 
\begin{align}
    \hat U[g] \, \ket{\psi_{B, V, J}} = e^{i \int d^D x \, \theta^a( g(x)) \hat G^a(x) } \, \ket{\psi_{B, V, J}} = \ket{\psi_{B, V, J}} \hspace{1cm} \forall g(x) 
   \label{eq:local_symmetry} 
\end{align}
where  $\theta_a(g(x))$ are dimensionless coordinates on the group manifold associated to the group element $g(x)$. To do this, note that \ref{eq:charge_operator} implies that $\hat a^\dagger_\alpha(x)$ and $\hat b^\dagger_\alpha(x)$ respectively transform with respect to right fundamental and left anti-fundamental of $\hat U[g]$ so that 
\begin{align}
	\hat U[g] \, \hat a^\dagger_\alpha(x) \, \hat U^\dagger[g]  = \hat a^\dagger_\beta(x)  D^{j_f}_{\beta \alpha}(g)
\end{align}
and
\begin{align}
	\hat U[g] \, \hat b^\dagger_\alpha(x) \, \hat U^\dagger[g]  =D^{j_f}_{\alpha \beta}(g^{-1}) \hat b^\dagger_\beta(x)  
\end{align}
where 
\begin{align}
	D^j_{\alpha\beta}(g(x)) = e^{i \theta_a(g(x)) T^{j \, a} }|_{\alpha \beta}
\end{align}
are the Wigner matrices for a representation $j$. Similarly, since $\hat A_\mu^a(x)$ lives in the Lie algebra (i.e. adjoint representation) of the group we find:  
\begin{align}
    &\hat U[g] \, \hat A_\mu^a(x) \, \hat U^\dagger[g] \\
    &\hspace{0.5cm}= \hat A_\mu^a(x) + \frac{1}{g} \hat D_\mu \theta^a(g(x))\\    
    &\hspace{0.5cm}= \hat A_\mu^a(x) + \frac{1}{g} \partial_\mu \theta^a(g(x)) + f^{abc} \hat A_\mu^b(x) \theta^c(g(x))
    \label{eq:transformation_A}
\end{align}
where we used the fact that the group manifold coordinates $\theta^a(g(x))$ also transform like the adjoint representation. 

Finally, since both the Fock vacuum $\ket{0}$ and the electric vacuum $\ket{s_E}$ transform trivially we get
\begin{align}
    \hat U[g]\ket{\psi_{B, V, J}}  = \int \mathcal{D}^2 \chi \, B[\chi_{\partial \mathcal{M} }] e^{ -\int_{\mathcal{M}} d^D x \  \hat{\mathcal{L}}[\hat{\tilde{A}}^a_\mu ,\chi,  \hat a^\dagger D[g] , D[g^{-1}] \hat b^\dagger    ]  \,   }  \ket{ 0 } \ket{s_E} \nonumber 
\end{align}
where $\hat{\tilde A}_\mu^a(x)$ is the right-hand side of \ref{eq:transformation_A}. Finally, a change of integration variables into leads to:
\begin{align}
    &\hat U[g] \ket{\psi_{B, V, J}} = \int \mathcal{D}^2 \chi \, B[D[g^{-1} ]\chi_{\partial \mathcal{M} }] 
      e^{ -\int_{\mathcal{M}} d^D x \  \hat{\mathcal{L}}[\tilde{\hat A}^a_\mu , D[g^{-1}] \tilde \chi, \hat a^\dagger D[g] , D[g^{-1}] \hat b^\dagger ]  \,    } \ket{\tilde \phi} \ket{s_E} \nonumber
\end{align} \ \ .

As promised, the symmetry/Gauss law is therefore present whenever 
\begin{align}
    \hat{\mathcal{L}}[\hat A^a_\mu ,\chi, \phi] = \hat{\mathcal{L}}[\tilde{\hat A}^a_\mu , D[g^{-1}]  \chi, \hat a^\dagger D[g] , D[g^{-1}] \hat b^\dagger ] \label{eq:gauge_invariance_Lagrangian}
\end{align}
and 
\begin{align}
	B[\chi_{\partial \mathcal{M} }]  = B[D[g^{-1} ]\chi_{\partial \mathcal{M} }] 
\end{align} 
i.e., when the generating Lagrangian and boundary functional are both singlets with respect to the group G. 

\subsection{Operator representation of the virtual spaces}
To remain closer in spirit to other continuous Tensor network methods, and especially those in one dimension, we will now introduce an alternative formulation of the gauge invariant CPEPS where the virtual degrees of freedom are explicitly represented as operators acting on a virtual Hilbert space. 

To achieve this we will pick a preferred direction $x_0$ and temporarily discretise space along it, with $a$ denoting the lattice spacing. Next, we define a $D-1$ dimensional Hilbert space with the associated operators $\hat \chi^j_\alpha(x_\perp)$, $\hat \chi^{\dagger \, j}_\alpha(x_\perp) $ and corresponding canonical momenta $\hat \pi^j_\alpha(x_\perp)$, $\hat \pi^{\dagger \, j}_\alpha(x_\perp) $ (note that we used small $\pi$ to differentiate from the physical canonical momenta). On this Hilbert space we define the transfer matrix 
\begin{align}
	T_i = e^{-a \int d^{D-1}x_{\perp} H[ \hat \chi(x_\perp), \hat \pi(x_\perp) , \hat A^a(x_0, x_\perp), a^\dagger(x_0, x_\perp), b^\dagger(x_0, x_\perp)  ]  } \ \ ,
\end{align}
so that
\begin{align}
	\braket{\chi_i |  T_i | \chi_{i+1}} = e^{-a \int d^{D-1}x_{\perp}  \hat{\mathcal{L}}[\hat A^a_\mu ,\chi, \hat a^\dagger, \hat b^\dagger] } \ \ , \label{eq:def_virtual_Hamiltonian}
\end{align}
in which $\ket{\chi_i}$ are $D-1$ dimensional states which are eigenvectors of the $\hat \chi^j_\alpha(x_\perp)$ and $\hat \chi^{\dagger \, j}_\alpha(x_\perp) $ operators. Using these definitions one can re-express the state as:
\begin{align}
	\ket{\psi_{B, V, J}} = tr_{aux}\del{ \hat{B}  \mathcal{P}e^{ -\int dx_0 \int d^{D-1}x_\perp \hat H[ \hat \chi(x_\perp), \hat \pi(x_\perp) , \hat A^a(x_0, x_\perp), a^\dagger(x_0, x_\perp), b^\dagger(x_0, x_\perp)  ]   }  } \ket{0} \ket{s_E} \label{eq:operator_formalism}
\end{align} 
where $\mathcal{P}$ is the path ordered exponential. For a generic $x_0$ dependent operator $\hat O(x_0)$ it is defined so that $\mathcal{P}e^{\int_{x_{start}}^{x_{end}} dx_0  \hat O(x_0)} = \lim_{N \rightarrow \infty} \prod_{n \in [1 \cdots N]} e^{\del{ \frac{x_{end}-x_{start}}{N} \hat O\del{x_{start} + n \frac{x_{end}-x_{start}}{N} }}}$ .

When the generating Lagrangian depends only on $\chi(x)$ and $\bar \chi(x)$ and not on their derivatives, one can check that 
\begin{align}
	\hat H &= \int d^{D-1} x_\perp \del{ \hat \pi^j_\alpha(x_\perp) \hat \pi^j_\alpha(x_\perp) - ig \hat A^a_{0}(x_0, x_\perp)  \hat J^a_{0}(x_\perp)  } \nonumber \\
		& \hspace{1cm} +\int d^{D-1} x_\perp \del{ ( \hat D^{(x_0)}_{x_\perp} \hat \chi^j_\alpha(x_\perp)  )^\dagger ( \hat D^{(x_0)}_{x_\perp} \hat \chi^j_\alpha(x_\perp)  )  } \nonumber \\
		& \hspace{1cm} +\int d^{D-1} x_\perp \del{ V[\hat \chi(x_\perp)] + J[\hat \chi(x_\perp), \hat a^\dagger(x_0, x_\perp), \hat b^\dagger(x_0, x_\perp)  ]  } \label{eq:H_virtual}
\end{align}
with
\begin{align}
	\hat J^a_{0}(x_\perp) =   -i \hat \pi^j_\alpha(x_\perp) T^{j \, a}_{\alpha \beta} \hat \chi^j_\beta(x_\perp)  + h.c.  
\end{align}
and
\begin{align}
	\hat D^{(x_0)}_{x_\perp} \hat \chi^j_\alpha(x_\perp) = \partial_{x_\perp} \hat \chi^j_\alpha(x_\perp) - i g \hat A^a_{x_\perp} (x_0, x_\perp) T^j_{\alpha \beta} \hat \chi^j_\beta(x_\perp)
\end{align}
is the correct choice for the virtual Hamiltonian. To verify this, one first inserts a complete set of $\hat \pi^j_\alpha(x_\perp)$ and $\hat \pi^{\dagger \, j}_\alpha(x_\perp) $  eigenstates into the left hand side of \ref{eq:def_virtual_Hamiltonian}. Upon performing the resulting Gaussian integrals the right hand side is immediately recovered. Note that the assumption that the potential and sources do not depend on derivatives of the virtual field can also be relaxed as is discussed in \cite{Peskin_an_introduction_to_quantum_field_theory}. One can easily check that the condition of gauge invariance implies that $\hat J^a_{0}(x_\perp)$ commutes $\hat \pi^j_\alpha(x_\perp) \hat \pi^j_\alpha(x_\perp)$, $( \hat D^{(x_0)}_{x_\perp} \hat \chi^j_\alpha(x_\perp)  )^\dagger ( \hat D^{(x_0)}_{x_\perp} \hat \chi^j_\alpha(x_\perp)  ) $ and $V[\hat \chi(x_\perp)] $. 

At this point, it is worthwhile to gather some intuition about these states. To do this, we recall that  $\exp\del{-i g \int_{x_{0 \, start}}^{x_{0 \, end}} dx_0 \hat A^a_0(x_0, x_\perp) T^a }$ is the creation operator of a quantum of electric flux in the $x_0$ direction. With this in mind, it is clear that the term $- ig \hat A^a_{0}(x_0, x_\perp)  \hat J^a_{0}(x_\perp)$ in \ref{eq:H_virtual} raises the electric field of the input $\ket{s_E}$ by exactly the flux of the virtual degrees of freedom. Since our state is Lorentz invariant the same argument holds for the other components of the electric field. 

\subsection{Truncation of virtual Hilbert space in one dimension}
In one spatial dimension, the virtual Hamiltonian becomes that of a zero dimensional quantum system  
\begin{align}
    \hat H = \hat\pi^j_\alpha \hat\pi^{\dagger \, j}_\alpha &- ig \hat A^a_{x_0}(x_0) (-i)\del{\hat \pi^j_\alpha T^{j \, a}_{\alpha \beta} \hat \chi^j_\beta - h.c. } +  V[\hat \chi]  + J[\hat \chi, \hat a^\dagger_\alpha(x_0), \hat b^\dagger_\alpha(x_0)]
\end{align} \
and we will now see that such states can be simplified by truncating the virtual Hilbert space at some finite occupation of the virtual degrees of freedom.

To explicitly achieve this truncation, we introduce virtual creation and annihilator operators
\begin{align}
	\hat{a}^{j \, P}_\alpha &= \frac{1}{\sqrt{2}}\del{ \sqrt{\Lambda} \hat \chi^j_\alpha + i \frac{1}{\sqrt{\Lambda}}\hat \pi^{j \,  \dagger}_\alpha } \\
	\hat{b}^{j \, P \, \dagger}_\alpha &= \frac{1}{\sqrt{2}}\del{\sqrt{\Lambda}\hat \chi^j_\alpha - i \frac{1}{\sqrt{\Lambda}} \hat \pi^{j \,\dagger}_\alpha  } . 
\end{align}
and the associated basis set 
\begin{align}
	\ket{   \{ \alpha_n, j_n, \alpha_m, j_m  \} }  = \frac{1}{\sqrt{N}} \prod_n a^{j_n \, P \, \dagger}_{\alpha_n}  \prod_m b^{j_m \, P \, \dagger}_{\alpha_m}   \ket{0}
\end{align}
that can be straightforwardly truncated. 

Finally, upon defining the matrices
\begin{align}
	F_{     \{ \alpha_n, j_n, \alpha_m, j_m  \}       ,    \{ \alpha'_n, j'_n, \alpha'_m, j'_m  \}    } &= \braket{ \{ \alpha_n, j_n, \alpha_m, j_m  \} | \hat \pi^j_\alpha \hat \pi^{\dagger \, j}_\alpha + V[\hat \chi] |\{ \alpha'_n, j'_n, \alpha'_m, j'_m  \}}  \\
	Q^a_{     \{ \alpha_n, j_n, \alpha_m, j_m  \}       ,    \{ \alpha'_n, j'_n, \alpha'_m, j'_m  \}    }  &=  \braket{ \{ \alpha_n, j_n, \alpha_m, j_m  \} | (-i)\del{\hat \pi^j_\alpha T^{j \, a}_{\alpha \beta} \hat \chi^j_\beta - h.c. }  |\{ \alpha'_n, j'_n, \alpha'_m, j'_m  \}} \\
	J^{a^\dagger}_{     \{ \alpha_n, j_n, \alpha_m, j_m  \}       ,    \{ \alpha'_n, j'_n, \alpha'_m, j'_m  \}    }  &= \braket{ \{ \alpha_n, j_n, \alpha_m, j_m  \} | J^{a^\dagger}[\hat \chi]  |\{ \alpha'_n, j'_n, \alpha'_m, j'_m  \}} \\
	J^{b^\dagger}_{     \{ \alpha_n, j_n, \alpha_m, j_m  \}       ,    \{ \alpha'_n, j'_n, \alpha'_m, j'_m  \}    }  &= \braket{ \{ \alpha_n, j_n, \alpha_m, j_m  \} | J^{b^\dagger}[\hat \chi]  |\{ \alpha'_n, j'_n, \alpha'_m, j'_m  \}} 
\end{align} 
we arrive at the truncated gauged continuous matrix product state
\begin{align}
        \ket{\psi} = tr_{aux} \del{ B \,  \mathcal{P} \exp{\int dx \left(   F - ig \hat A^a_x(x) Q^a + J^{a^\dagger}_\alpha \, \hat a^\dagger_\alpha(x) + J^{b^\dagger}_\alpha \, \hat b^\dagger_\alpha(x)  \right)  } } \ket{s_E}\ket{0_{fock}}  \label{eq:gauged_CMPS}
\end{align} 
that is the natural generalisation of the CMPS proposed in \cite{Verstraete_2010}.

Note that this trunctation of the virtual Hilbert space naturally results in a truncation of the physical electric field as well as the physical particle density. Hence these states cannot access the full Hilbert space of the quantum field theory. In fact, this is a feature rather then a bug since it is well known that operators defined on the full quantum field theory Hilbert space are plagued with UV-divergences. 

\subsection{Extension to Fermions}
A priori, there is no reason against generalising gauged continuous tensor networks to states involving fermionic matter. In this case the virtual degrees of freedom should also be fermionic so that the integration over virtual field would become a Grassmann integral. As mentioned in \cite{Tilloy_2019} it may be more convenient to circumvent this peculiarity and immediately define such states in the operator formalism. 

\section{Computation of expectation values through generating functionals}
To compute observables with respect to these states, it is convenient to define generating functionals such as:
\begin{align}
	Z[j^{a^\dagger}, j^{b^\dagger}, j^A] = \frac{ \braket{ \psi_{B, V, J}|   e^{\int d^D x \del{ j^{a^\dagger}_\alpha(x) a^{j}_\alpha(x)   + j^{b^\dagger}_\alpha(x) b^{j}_\alpha(x)  -i g \, j^{A\, a}_\mu(x) A^a_\mu(x)    }}  | \psi_{B, V, J}  }}{ \braket{ \psi_{B, V, J} |  \psi_{B, V, J}} }
\end{align} 
which can be used to compute the expectation value of open Wilson lines (with charges as endpoints) through explicit expansion of the path ordered exponential that constitutes the Wilson line i.e.: 
\begin{align}
	&\frac{  \braket{ \psi_{B, V, J} |  \hat a^\dagger_\alpha(x) P e^{-i g \int_{path}  dp \, t_\mu(p) \hat A^a_\mu(p) T^a }|_{\alpha \beta} \hat b^\dagger_\beta(y) |  \psi_{B, V, J}} }{  \braket{ \psi_{B, V, J} |  \psi_{B, V, J}} } \nonumber \\
	& \hspace{1cm}= \frac{\delta Z[j^{a^\dagger}, j^{b^\dagger}, j^A]  }{\delta j^{a^\dagger}_\alpha(x) \, \delta j^{b^\dagger}_\alpha(y) } + \int_{path} dp \, t_\mu(p) \frac{\delta Z[j^{a^\dagger}, j^{b^\dagger}, j^A]  }{\delta j^{a^\dagger}_\alpha(x) \,  \delta j^{A \, a}_\mu(z) \, \delta j^{b^\dagger}_\alpha(y) } T^a_{\alpha \beta} + ...
\end{align} \ \ .

Similarly we define 
\begin{align}
Z[j^E] = \frac{ \braket{ \psi_{B, V, J} |   e^{\int d^D x \del { j^{E \, a}_\mu(x) E_\mu^a(x)}}   | \psi_{B, V, J}} }{ \braket{ \psi_{B, V, J} |  \psi_{B, V, J}} }
\end{align}
which allows for the computation of expectation values of electrical fields through taking functional derivatives. 

As mentioned in the introduction, in $D \leq 3 $ these generating functionals lead to finite correlation functions when the generating Lagrangian is Gaussian (note that this implies $g=0$). In what follows we will propose two methods to go beyond this. First we will outline how to express the generating functionals as a perturbative expansion in $g$. Generally this series must be renormalized through the introduction of a UV-cutoff or counterterms. Second we will outline a TN inspired contraction scheme that iteratively reduces the dimension of the problem to one that lives in one dimension. The final D=1 case can then be naturally regulated by truncating the virtual Hilbert space using the methods discussed in section 2.3.

\subsection{Gauged Continuous Gaussian tensor network states}
For simplicity we will focus on the $U(1)$ case, assume a single virtual field and assume translation symmetry in the thermodynamic limit, under these circumstances the most general gauged CPEPS that is Gaussian in the virtual fields is:
\begin{align}
	\ket{\psi_{B, V, J}} = \int D^2 \chi e^{-\int d^D x \del{  \del{\hat D_\mu \chi(x)}^\dagger  \hat D^\mu(x)  + m^2 \bar \chi(x) \chi(x) + J^{a^\dagger} \hat a^\dagger(x) \chi(x) + J^{b^\dagger} \bar \chi(x) \hat b^\dagger(x)   }} \ket{0} \ket{s_E}
\end{align} 
which can be conveniently rewritten in terms of coherent states as
\begin{align}
	\ket{\psi_{B, V, J}} = \int D^2 \chi e^{-\int d^D x \del{  \del{\hat D_\mu \chi(x)}^\dagger  \hat D^\mu(x)  + m^2 \bar \chi(x) \chi(x) }} \ket{ J^{a^\dagger} \chi(x) ;  J^{b^\dagger} \bar \chi(x)  } \ket{s_E}
\end{align} 
\ \ .

Using properties of coherent states, the generating functional can be expressed as
\begin{align}
	Z[j^{a^\dagger}, j^{b^\dagger}, j^A] &= \frac{1}{N} \int DA \int D^2 \chi' D^2 \chi e^{-1/2 \int d^D x  \del{ \boldsymbol{\chi}^T(x) (\sigma_x \otimes \boldsymbol\omega) \boldsymbol \chi(x)  + \boldsymbol \chi^T(x) \boldsymbol J(x) - i g \,  j^{A}_\mu(x) A_\mu(x) } }
\end{align}
where $N = \braket{\psi_{B,V,N} | \psi_{B,V,N}}$, 
\begin{align}
	\boldsymbol{\chi}(x) &= ( \chi'(x) , \, \chi(x) , \, \bar\chi'(x) , \, \bar\chi(x)  )^T  \\
	\boldsymbol{J}(x) &= (  j^{b^\dagger}(x) \bar J^{b^\dagger}  , \, 0 , \,   j^{a^\dagger}(x) \bar J^{a^\dagger} , \, 0)^T  \\
	\boldsymbol{\omega} &= \begin{bmatrix} (\partial_\mu + i g A_\mu(x))^2 - m^2 & \bar J^{b^\dagger} J^{b^\dagger} \\  \bar J^{a^\dagger} J^{a^\dagger} & (\partial_\mu + i g A_\mu(x))^2 - m^2 \end{bmatrix}
\end{align} 
and the integral over gauge fields configurations has arisen because we have inserted $1 = \int DA \ket{A} \bra{A}$ where $\ket{A}$ are eigenstates of the gauge field operators $A(x)$.

Since the integral over the virtual fields is Gaussian, we can easily get rid of it to find
\begin{align}
	Z[j^{a^\dagger}, j^{b^\dagger}, j^A] = \frac{1}{\tilde{N}} \int DA \frac{e^{ \int d^D x (-i g\,  j^{A}_\mu(x) A_\mu(x)) }}{ det( \boldsymbol\omega[A(x)]) } e^{-\frac{1}{2} \int d^D x d^Dy \boldsymbol J^T(x) (\sigma_x \otimes G(x-y)) \boldsymbol J(y) }
	\label{eq:generating_functional_particles}
\end{align}
where 
\begin{align}
	\boldsymbol \omega G(x-y) = \delta^D(x-y)
\end{align}
and $\tilde N$ is some new normalization that can be fixed by requiring that $Z=1$ when the sources are set to 0. Generally, $G(x-y)$ and $det(\boldsymbol \omega)$ will be polynomial functions of the gauge fields $A_\mu(x)$ so that the integration $\int DA$ is difficult to tackle. 

\subsubsection*{perturbative expansion of the generating functional}
Since
\begin{align}
	\boldsymbol \omega(x) = \boldsymbol \omega_0(x) + g \boldsymbol \omega_1(x) + g^2 \boldsymbol \omega_2(x)
\end{align}
with 
\begin{align}
	\boldsymbol \omega_0(x)  &= \begin{bmatrix} \partial^2 - m^2 & \bar J^{b^\dagger} J^{b^\dagger} \\  \bar J^{a^\dagger} J^{a^\dagger} & \partial^2 - m^2 \end{bmatrix} \\
	\boldsymbol \omega_1(x) &= \mathbb{1}_{2 \times 2 } \, (2 i A^\mu(x) \partial_\mu + i \partial_\mu A^\mu(x) ) \\ 
	\boldsymbol \omega_2(x) &= \mathbb{1}_{2 \times 2 } \, A_\mu(x) A^\mu(x)
\end{align}
it makes sense to suggest
\begin{align}
		G(x-y) = G_0(x-y)+ g G_1(x-y) + g^2 G_2(x-y) + ...
\end{align}
so that $\boldsymbol \omega G(x-y) = \delta^D(x-y)$ can be solved order by order. 

Doing so gives:
\begin{align}
	G_0(x-y) = \int d^D k e^{ik(x-y)} \begin{bmatrix} k^2 + m^2 & \bar J^{b^\dagger} J^{b^\dagger} \\  \bar J^{a^\dagger} J^{a^\dagger} & k^2 + m^2 \end{bmatrix}^{-1}
\end{align}
and 
\begin{align}
	G_1(x-y) = \int d^D z G_0(x-z) \omega_1(z) G_0(z-y)
\end{align}
etcetera. 

Similarly
\begin{align}
	det(\boldsymbol \omega) &= det\del{\boldsymbol\omega_0 + g \int d^D y \boldsymbol \omega_0 G_0(x-y) \boldsymbol \omega_1 + g^2 \int d^D y \boldsymbol \omega_0 G_0(x-y) \boldsymbol \omega_2} \\ 
	&=det( \boldsymbol \omega_0)e^{\sum_{n \geq 1} \frac{1}{n} tr\del{ g \int d^D y \boldsymbol G_0(x-y) \boldsymbol \omega_1 + g^2 \int d^D y \boldsymbol G_0(x-y) \boldsymbol \omega_2  }^n }
\end{align}
where we used $det(A) = \exp(\text{tr}(\text{log}(A)))$ and expanded the logarithm. 

Re-inserting all this into \ref{eq:generating_functional_particles} and expanding the exponents in $g$ results in an expression of the fom
\begin{align}
	Z[j^{a^\dagger}, j^{b^\dagger}, j^A]  = \frac{1}{\tilde{N}} \int DA e^{ \int d^D x (- i g j^{A}_\mu(x) A_\mu(x)) } f(A) e^{-\frac{1}{2} \int d^D x d^Dy \boldsymbol J^T(x) (\sigma_x \otimes G_0(x-y)) \boldsymbol J(y) }
\end{align}
for some polynomial $f(A)$. In principle these integrals can be solved order by order and used to optimize these states with respect to some Hamiltonian. 

\subsubsection*{Proposed Monte-Carlo sampling}
Alternatively one may expand $A_\mu^a(x)$ in terms of some truncated set of ortogonal functions and perform the integration using Monte-Carlo sampling. Such an algorithm would be very close in spirit to the ones already explored in \cite{Kelman_2024}.

\subsection{Beyond Gaussianity : the transfer matrix formalism}
In this section we, will build a transfer matrix formalism to compute the generating functional non-perturbatively. The upshot of this will be that the computation of certain physically relevant D-dimensional generating functionals reduces to the optimization of a D-1 dimensional eigenvalue problem for an effective operator. 

We wish to compute:
\begin{align}
 Z &= \frac{1}{N}\bra{0}\bra{s_E} tr_{aux'} \del{ \hat B \mathcal{P} e^{-\int dx_0 dx_\perp \hat H^{\dagger '} }  } \\
 &\hspace{2cm}e^{\int d^D x \del{ j^{a^\dagger}_\alpha(x) a^{j}_\alpha(x)   + j^{b^\dagger}_\alpha(x) b^{j}_\alpha(x)  -i g \, j^{A\, a}_\mu(x) A^a_\mu(x)    }} \nonumber \\
 &\hspace{3cm} tr_{aux} \del{ \hat B  \mathcal{P}  e^{-\int dx_0 dx_\perp \hat H} } \ket{0} \ket{s_E} \nonumber
\end{align}
where $\hat H$ is given by \ref{eq:H_virtual} and H' is the same but acting on the virtual degrees of freedom associated to the bra instead of the ket.

To proceed, one should temporarily discretise the $x_0$ direction with lattice spacing $a$ and rewrite the path ordered exponential as a product over discretised $x_0$. Then one commutes $H'$ all the way to the right while carefully applying the Baker Campbell Haussdorf formula. Note that the term arising from the commutator contains terms such as $a^2 [a^\dagger(x_0, x_\perp), a(x_0, x_\perp)] = a \delta(x_\perp - y_\perp)$ that do not vanish in the continuum. Finally, one merges all remaining terms into one exponential, here one finds commutator terms such as $a^2 [\chi(x_\perp), \pi(y_\perp)]$ that vanish in the continuum. After all this one finds: 
\begin{align}
	Z &= \frac{1}{N} \bra{s_E} tr_{aux, aux'} \Big( B B'  \mathcal{P}  e^{-\int dx_0 dx_\perp \del{  \boldsymbol H  } }         \Big) e^{-i g \int dx_0 dx_\perp j^{A a}_\mu (x) \hat A^a_\mu(x) }   \ket{s_E}
\end{align}
where 
\begin{align}
	\boldsymbol H = H + H' &+ \bar J^{a^\dagger} \hat \chi^{j \, \dagger \, '}_\alpha(x_\perp)(J^{a^\dagger}\chi^{j }_\alpha(x_\perp) + j^{a^\dagger}_\alpha(x_0, x_\perp)    ) \nonumber \\
	&+ \bar J^{b^\dagger} \hat \chi^{j \, '}_\alpha(x_\perp)(J^{b^\dagger}\chi^{j \, \dagger}_\alpha(x_\perp) + j^{b^\dagger}_\alpha(x_0, x_\perp)    ) 
\end{align}
is an operator that acts on the virtual degrees of freedom from the bra and ket $\hat \chi(x_\perp)$, $\hat \pi(x_\perp)$, $\hat \chi'(x_\perp)$, $\hat \pi'(x_\perp)$   as well as the gauge degrees of freedom $\hat A^a_\mu(x_0, x_\perp)$. The 3th and 4th terms in $\boldsymbol H$ arise from careful application of Baker Campbell Haussdorf as discussed earlier.

Next, we extract the terms containing $\hat J_0(x_\perp)\hat A_0(x_0, x_\perp)$and insert $1 = \int DA_0 \ket{A_0} \bra{A_0}$ into the expression. We find: 
\begin{align}
	Z &= \frac{1}{N} \bra{s_{E_\perp}} tr_{aux, aux'} \Big( B B'  \mathcal{P}  e^{-\int dx_0 dx_\perp \del{  \boldsymbol H_\perp  } }         \Big) \nonumber \\
	&\hspace{1.5cm}\int DA_0 e^{-i g \int dx_0 dx_\perp  A_0(x_0, x_\perp) \del{j_0^A + \hat J_0(x_\perp) - \hat J'_0(x_\perp) } } \nonumber\\
	&\hspace{3cm}e^{-i g \int dx_0 dx_\perp j^{A a}_\perp  (x) \hat A^a_\perp(x) }   \ket{s_{E_\perp}}
\end{align}
where $\ket{S_{E_\perp}}$ is the electrical singlet in the remaining $D-1$ dimensions and $\boldsymbol{H}_\perp$ is the part of $\boldsymbol{H}$ that does not contain $A_0$. Where we observe that, for a fixed $x_0$, the second line of this equation is simply a non-normalized projector onto the subspace of the virtual Hilbert space (at that time) where $j_0^A(x_\perp) = \hat J'_0(x_\perp) - \hat J_0(x_\perp)$ is satisfied. 

Finally, to actually reduce the dimensionality of the problem we focus on sources $j^{a^\dagger}_\alpha(x_0, x_\perp)$, $j^{b^\dagger}_\alpha(x_0, x_\perp)$ and $j_\mu^{A \, a}(x_0, x_\perp)$ that are only nonzero when $x_0 = 0$. With this assumption, we can rewrite the generating functional as the D-1 dimensional expectation value: 
\begin{align}
	Z &= \frac{1}{\tilde{N}} \bra{s_{E_\perp}} \bra{v} e^{- \int d x_\perp \boldsymbol{H}_\perp }e^{-i g \int dx_\perp j^{A a}_\perp  (x_\perp) \hat A^a_\perp(x_\perp) } \ket{v}   \ket{s_{E_\perp}} \label{eq:effective_functional}
\end{align}
where $\ket{v}$ is the D-1 dimensional state on the Hilbert space of virtual degrees of freedom from the bra and ket that optimizes
\begin{align}
	\lambda =  \bra{s_{E_\perp}} \bra{v_\lambda} e^{- \int d x_\perp \boldsymbol{H}_\perp|_{j = 0} } \ket{v_\lambda}   \ket{s_{E_\perp}}\label{eq:boundary_state}
\end{align} 
under the constraint that $\del{ \hat J'_0(x_\perp) - \hat J_0(x_\perp) } \ket{v} = 0$, such as state can easily be constructed as a CPEPS of the form
\begin{align}
	\ket{v_\lambda} = \int D^2 \chi_{TM} B[\chi_{TM}] e^{\int_{\mathcal{M}_{D-1}}d^{D-1}x_\perp \hat{\mathcal{L}}[\hat A^a_\perp, \chi_{TM}, a^\dagger a^{\dagger' } , b^\dagger b^{\dagger'}]    } \ket{0}
\end{align} 
that contains only products of the creation operators of virtual particles living in the bra an ket Hilbert spaces.

Finally, to truly make this argument iteratively we note that \ref{eq:effective_functional} and \ref{eq:boundary_state} can both be further reduced in dimension by picking another preferred direction and introducing a basis of identities along that direction. Upon doing this the integration over the gauge fields in this new preferred direction will reduce to a projector and one makes a new fix point of the transfer matrix that properly accounts for this contraint. Eventually one ends up with a normal CMPS construction and as mentioned before, here the problem can finally be nuturally regularized by truncating the dimension of the virtual Hilbert space.

\section{Gauged CPEPS as continuum limits of gauged PEPS}
Let us now demonstrate that gauged CPEPS are the continuum limit of suitable sequences of gauged PEPS. This is particular relevant in light of the current proof that gauged PEPS are the most general gauge invariant states on the lattice \cite{Kull_Classification_of_matrix_product_states_with_a_local_gauge_symmetry,Blanik_Internal_structure_gauge_invariant_PEPS}. Indeed, with this in mind it seems to be so that gauged CPEPS are the most general gauge invariant states in the continuum. To achieve this goal we will first set the stage be reviewing the usual construction in a language that suits our needs. 

First we introduce PEPS with global symmetries in second quantization rather than the usual first quantized approach. Second, we will review the gauging procedure in second quantization and write down the resulting PEPS with local gauge symmetry. Finally, we take the continuum limit to recover the above proposed gauged CPEPS.

\subsection{PEPS with a global symmetry}
It is well known that the definition of PEPS requires the introduction of some auxiliary Hilbert spaces on the bonds of the lattice \cite{Cirac_2021}. Since we eventually plan on taking a continuum limit, the dimension of each of these spaces ought to be infinite. The reason behind this is explained in detail in \cite{Tilloy_2019}, a less rigorous argument is that finite dimensionful correlation lengths $\chi_{corr}[m] = \chi_{latt} a[m]$ imply infinite dimensionless correlation as the lattice spacing $a$ approaches that of the continuum. Therefore, the underlying PEPS should have infinite bond dimension as it approaches the continuum PEPS. With this in mind, it is natural to consider them to be associated to virtual fields in second quantization rather than Hilbert spaces in first quantization. Furthermore, since we will want to PEPS to have a global symmetry with respect to the group G, it is natural that these virtual fields carry irrep labels $j$ and color labels $\alpha$. 

With all this in mind we define operators $\hat \chi_{\alpha \, \bfn \,  \bfei}^{j}$ and $\hat \pi_{\chi \ \alpha \, \bfn \,  \bfei}^{j}$  for each position $\bfn$ that live on the links pointing away from $\bfn$ in the positive $\bfei$ direction. Similarly, we define $\hat \eta_{\alpha \, \bfn \,  \bfei}^{j} $ and $\hat \pi_{\eta \ \alpha \, \bfn \,  \bfei}^{j}$ for the links pointing towards negative $\bfei$ direction. Finally, we define the discrete matter field $\hat \phi_{\alpha \, \bfn}$ and its canonically conjugate momentum $\hat \pi_{\phi \ \alpha \,  \bfn} $. As before we introduce bases $\ket{\phi_\bfn} = \prod_\alpha \ket{\phi_\bfn^\alpha}$, $\ket{\chi_{\bfn \, \bfei}} = \prod_{j \ \alpha} \ket{\chi_{\bfn \ \bfei}^{j \ \alpha}  }$ and $\ket{\eta_{\bfn \, \bfei}} = \prod_{j \ \alpha } \ket{\eta_{\bfn \ \bfei}^{j \ \alpha}  }$ of eigenvectors for the respective field operators. For clarity, we depict the various fields operators and their locations on the lattice below.
\begin{figure}
    \centering
    \includegraphics[width=8cm]{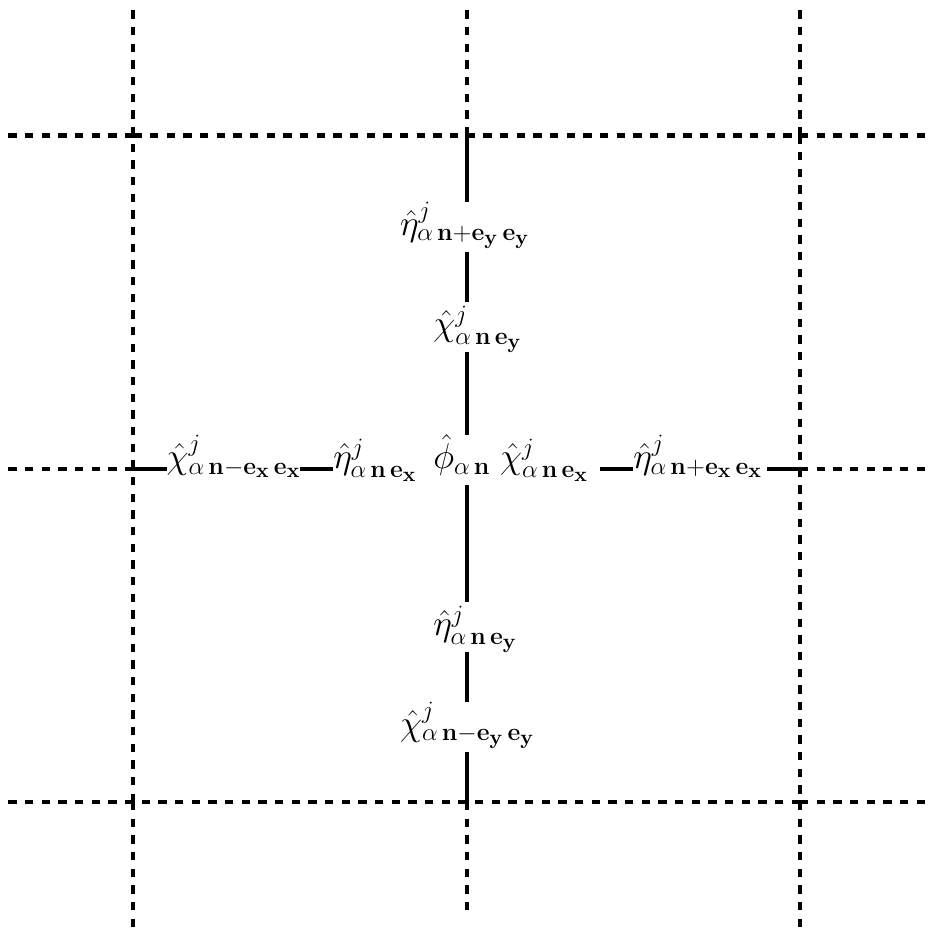}
\end{figure}

On these spaces, we define the group elements 
\begin{align}
    \hat \theta(g)_\bfn = e^{i \phi_a(g) \hat Q^a_{\phi \, \bfn} } 
\end{align}    
where
\begin{align}
    Q^a_{\phi \, \bfn} = -i \del{ \hat\pi_{\phi \ \alpha \, \bfn} \, T_{\alpha \beta}^a \, \hat \phi_{\beta \, \bfn \, \bfei}^{\beta} + h.c.  }  \ \ \ ,
\end{align}    
the action of these group elements on the field operators and their eigenstates is similar to what we discussed before i.e. anti-fundamental. 

Now, define the PEPS 
\begin{align}
    \ket{\psi} = \otimes_{\bfn \, \bfei} \bra{L_{\bfn \, \bfei}}  \otimes_{\bfn} \ket{A_\bfn}
\end{align}
from the on site tensors
\begin{align}
    \ket{A_\bfn} = \int \mathcal{D}^2\phi_\bfn \mathcal{D}^2\eta_\bfn \mathcal{D}^2\chi_\bfn  \ A(\phi_\bfn, \eta_\bfn, \chi_\bfn) \ket{\phi_\bfn} \ket{\eta_\bfn} \ket{\chi_\bfn} \ \ ,
\end{align}
and bond tensors 
\begin{align}
    \ket{L_{\bfn \ \bfei} } = \int  \mathcal{D}^2\chi \mathcal{D}^2\eta \ \delta^2(\chi_{\bfn \ \bfei   } - \bar{\eta}_{\bfn + \bfei \ -\bfei}  )  \ \ \ .
\end{align}
With these, the contracted PEPS is: 
\begin{align}
    \ket{\psi} = \int \mathcal{D}^2\phi \mathcal{D}^2\chi \prod_{\bfn} A(\phi_\bfn, \bar\chi_{\bfn - \bfei \, \bfei} , \chi_{\bfn \, \bfei}) \ket{\phi_\bfn} 
    \label{eq:PEPS_global_symmetry}
\end{align}
and it easy to see that this is symmetric with respect to the action of the the global symmetry $\hat U(g) = \otimes_\bfn \hat \theta(g)_\bfn$ if 
\begin{align}
    A \left(\phi_\bfn, \eta_\bfn, \chi_\bfn \right) = A(D(g^{-1}) \phi_\bfn, \eta_\bfn D(g^{-1}), D(g) \chi_\bfn )   \ \ \ , \label{eq:tensor_symmetry}
\end{align}
indeed, to see this one simply has to do a change of integration variables as before. 

Note that \ref{eq:tensor_symmetry} also implies that
\begin{align}
    \theta(g)_\bfn \prod_\bfei \theta_\eta(g)_{\bfn \bfei} \, \theta_\chi(g)_{\bfn \bfei} \ket{A_\bfn} = \ket{A_\bfn} \label{eq:virtual_symmetry}
\end{align}
and similarly, due to their definition, the bond states satisfy 
\begin{align}
    \theta_\chi(g)_{\bfn \, \bfei} \theta_\eta(g)_{\bfn+\bfei \, \bfei} \ket{L_{\bfn \, \bfei}} = \ket{L_{\bfn \, \bfei}} \label{eq:bond_symmetry}
\end{align}
where we defined the group elements acting on the virtual spaces as
\begin{align}
    \hat \theta_\eta(g)_{\bfn \, \bfei} = e^{i \phi_a(g) \hat Q^a_{\eta \, \bfn \, \bfei}   } \\
    \hat \theta_\chi(g)_{\bfn \, \bfei} = e^{-i \phi_a(g) \hat Q^a_{\chi \, \bfn \, \bfei}  }
\end{align}
where
\begin{align}
    \hat Q_{\eta \, \bfn \, \bfei}^a =  -i \del{ \hat \pi_{\eta \ \alpha \,\bfn \bfei}^{j } T_{\beta \alpha}^{j \, a} \hat \eta _{\beta \, \bfn \bfei}^{j} + h.c. } \\
    \hat Q_{\chi \, \bfn \, \bfei}^a =  -i \del{ \hat \pi_{\chi \ \alpha \bfn \, \bfei}^{j} T_{\alpha \beta}^{j \, a} \hat \chi _{\beta \, \bfn \bfei}^{j} + h.c. }   
\end{align}
so that 
\begin{align}
    \hat \theta_\eta(g)_{\bfn \bfei} \hat \eta^j_{\alpha \bfn \bfei} \hat \theta^\dagger_\eta(g)_{\bfn \bfei} = \eta^j_{\beta \bfn \bfei} D(g^{-1})_{\beta \alpha} 
\end{align}   
and
\begin{align}
    \hat \theta_\chi(g)_{\bfn \bfei} \hat \chi^j_{\alpha \bfn \bfei} \hat \theta^\dagger_\chi(g)_{\bfn \bfei} = D(g)_{\alpha \beta} \chi^j_{\beta \bfn \bfei}  \ \ \ .
\end{align}

\subsection{From global symmetry to local symmetry}
One can lift this global symmetry to a local one by minimally coupling the PEPS to gauge degrees of freedom that reside on the links of the lattice. To start, we define bosonic operators $\hat L^a_{\bfn \bfei}$ and $\hat R^a_{\bfn \bfei}$ that act on the link pointing away from site $\bfn$ in the $\bfei$ direction. These commute across different links and on the same link they satisfy 
\begin{align}
    [\hat L^a_{\bfn \bfei} , \hat L^b_{\bfn \bfei}] &= -i f^{abc} L^c_{\bfn \bfei}  \\
    [\hat R^a_{\bfn \bfei} , \hat R^b_{\bfn \bfei}] &= \phantom{-}i f^{abc} R^c_{\bfn \bfei} \ \ \ .
\end{align}
Using these, we define left and right group elements 
\begin{align}
    \hat \theta_L(g)_{\bfn \bfei} &= e^{-i \phi_a(g) \hat L^a_{\bfn \bfei} } \\
    \theta_R(g)_{\bfn \bfei} &= e^{+i \phi_a(g) \hat R^a_{\bfn \bfei}}  
\end{align}
and the basis $\ket{g_{\bfn \bfei}} \, \forall \, g \, \in \, G$ so that 
\begin{align}
    \hat \theta_L(g_L)_{\bfn \bfei} \hat \theta_R(g_R)_{\bfn \bfei} \ket{g}_{\bfn \bfei} = \ket{g_L g g_R^{-1}}_{\bfn \bfei} \ \ \ .
\end{align}
Another useful basis is the electric field basis 
\begin{align}
    \ket{j m n}_{\bfn \bfei}  = \int dg \sqrt{\frac{|G|}{dim(j)}} D_{m n}^j(g) \ket{g}_{\bfn \bfei}    
\end{align}
where 
\begin{align}
    &\hat\theta_L(g_L) \hat\theta_R(g_R) \ket{j n m}_{\bfn \bfei} \nonumber\\
    &\hspace{1cm}= D^j_{m m'}(g_L^{-1})\ket{j m' n'}_{\bfn \bfei} D_{n' n}^j(g_R) \ \ \ .   
\end{align}
The singlet state $\ket{j = 0}$ corresponds to the zero electric field ket and will be denoted as $\ket{s_E}$. An in depth review of the relevant Hilbert space can be found in \cite{roose2025gaugingsuperpositionfermionicgaussian}. 

Using these bases we are now ready to define
\begin{align}
    \hat U_{\text{Gauge} \, \bfn \bfei} = \int dg \  e^{i \theta_a(g) \hat Q_{\eta \bfn \bfei}^a  }  \ket{g}_{\bfn \bfei} \bra{g}_{\bfn \bfei} 
\end{align}
which is a controlled raising operator for the electrical field on the link $\bfn \, \bfei$. Note, we can also interpret this operator as a controlled left group action on the $\eta$ variables, this will be helpful later. Crucially, using change of integration variables, one can show that this operator satisfies:
\begin{align}
    \theta_L(g)_{\bfn \, \bfei} \, U_{\text{Gauge} \, \bfn \bfei} \ket{s_E} &= \theta_\eta(g^{-1})_{\bfn \bfei} \, U_{\text{Gauge} \, \bfn \bfei} \ket{s_E} \nonumber \\
    \theta_R(g)_{\bfn \, \bfei} \, U_{\text{Gauge} \, \bfn \bfei} \ket{s_E} &= U_{\text{Gauge} \, \bfn \bfei} \, \theta_\eta(g)_{\bfn \bfei} \ket{s_E} \label{eq:props_gauger2}
\end{align}
and with that knowledge one can check that the gauged PEPS 
\begin{align}
    \ket{\psi_\text{gauged}} = \otimes_{\bfn \, \bfei} \bra{L_{\bfn \, \bfei}} \otimes_{\bfn \, \bfei} U_{\text{gauge} \, \bfn \bfei  }  \otimes_{\bfn} \ket{A_\bfn}
\end{align}
has a local symmetry with respect to the action of $\theta(g)_\bfn \prod_\bfei \theta_L(g)_{\bfn \bfei} \theta_R(g)_{\bfn -\bfei} $. The full derivation can be found in \cite{Kelman_2024}, here we simply note that \ref{eq:props_gauger2} together with \ref{eq:bond_symmetry} lift the local symmetry of the tensors \ref{eq:virtual_symmetry} to a physical symmetry of the state.

Finally, explicitly contracting the gauged PEPS leads to : 
    \begin{align}
        \ket{\psi_\text{gauged}} = \int \mathcal{D}^2\phi \mathcal{D}^2\chi dg  \prod_\bfn A( \phi_\bfn, \bar \chi_{\alpha \ \bfn-\bfei \, \bfei}^{j} D(g^{-1}_{\bfn-\bfei \, \bfei})^j_{\alpha \beta},  \chi^{j}_{\beta \ \bfn \, \bfei}  ) \ket{g}_{\bfn \, \bfei} \ket{\phi_\bfn}
    \end{align}
where we used the fact that $U_{Gauge}$ can be interpreted as a controlled rotation on the virtual degrees of freedom. 

\subsection{Taking the continuum limit}
To take the continuum limit, let us first assume that 
\begin{align}
    A(\phi, \chi, \eta) = e^{\mathcal{A(\phi, \chi, \eta)}} \label{eq:exponential_tensor}
\end{align}
so that the contracted state becomes: 

\begin{align}
        \ket{\psi_\text{gauged}} = \int \mathcal{D}^2\phi \mathcal{D}^2\chi dg  e^{\sum_\bfn \mathcal{A}\left( \phi_\bfn, \bar \chi_{\alpha \ \bfn-\bfei \, \bfei}^{j} D(g^{-1}_{\bfn-\bfei \, \bfei})^j_{\alpha \beta},  \chi^{j}_{\beta \ \bfn \, \bfei}  \right)} \prod_{\bfn \bfei}\ket{g}_{\bfn \, \bfei} \ket{\phi_\bfn} 
    \end{align}    
where without loss of generality we split $\mathcal{A}$ into its purely virtual part and the rest i.e.:
$\mathcal{A}(\phi, \eta, \chi) = K(\eta, \chi) + V(\phi, \eta, \chi )$. Following \cite{Tilloy_2019} we assume: 
\begin{align}
    K(\chi, \eta) = \frac{-Z_0}{2}\sum_\bfei \del{ \bar \chi_{\bfn \, \bfei} - \eta_{\bfn \, \bfei} }\del{ \chi_{\bfn \, \bfei} - \bar \eta_{\bfn \, \bfei} } \ \ \ .
\end{align}

To go to the continuum, we must replace dimensionless lattice fields with dimensionful continuum fields. In particular, we define $\phi(x = \bfn a) = a^{[\phi(x)]} \phi_{\bfn }$ and $\chi(x = (\bfn + \bfei/2)a ) = a^{[\chi(x)]} \chi_{\bfn \, \bfei}$ where $[\phi(x)] = [\chi(x)] = \frac{D-1}{2}$ for the scalar fields and $A_i^a(x = (\bfn + \bfei/2)a ) = \frac{1}{ag_{QFT}} \phi^a(g_{\bfn \, \bf e_i}) $ for the gauge manifold coordinates. With this we get: 
\begin{align}
    &\chi^j_{\alpha \ \bfn \, \bfei} - D(g_{\bfn-\bfei \, \bfei})_{\alpha \beta} \chi^j_{\beta \ \bfn - \bfei \, \bfei} = \frac{a}{a^{[\chi(x)]}} D_i \chi^j_\beta(\bfn)
\end{align}
where we assumed $a\rightarrow0$ and remembered the definition of the covariant derivative \ref{eq:covariant_derivative}. From this we immediately get: 
\begin{align}
    &K( \bar \chi_{\alpha \ \bfn-\bfei \, \bfei}^{j} D(g^{-1}_{\bfn-\bfei \, \bfei})^j_{\alpha \beta},  \chi^{j}_{\beta \ \bfn \, \bfei}  ) \nonumber\\
    &\hspace{3cm}= \frac{-Z_0}{2}\frac{a^2}{a^{2[\chi]}}  \bar D_i \bar \chi^j_\beta(\bfn) D_i \chi^j_\beta(\bfn)
\end{align}
so that the continuum limit of the gauged PEPS will be : 
    \begin{align}
        \ket{\psi_\text{gauged}} = \int \mathcal{D}^2\phi \mathcal{D}^2\chi dA  e^{-\int d^D x \bar D_i \bar\chi(x) D^i \chi(x) + V(\phi(x), \chi(x) )  }\ket{A(x)}\ket{\phi(x)} \ .
    \end{align}

Finally we can bring this into the proposed form \ref{eq:gauged_CPEPS}  by using $1 = \int DA \ket{A}\bra{A}$ and $\braket{A|s_E} = 1$ as well as $1 = \int D^2 \phi \ket{\phi}\bra{\phi}$ and $\braket{\phi|0_{fock}}$. Indeed, after inserting \ref{eq:phys_particles} and normal ordering with respect to the Fock vacuum one recovers \ref{eq:gauged_CPEPS} with some suitable $\tilde{V}$.

\section{Outlook and conclusion}
In this work, we proposed a new class of manifestly gauge invariant states in the continuum and demonstrated that they are the continuum limit of sequences of gauged PEPS. Furthermore, we demonstrated that in one dimension these states reduce to gauged CMPS. The intuition behind these states is that they take conventional globally symmetric continuous projected entangled pair states and lift the global symmetry to a local one. This is possible because tensor networks with global symmetries already respect a virtual Gauss law. 

Since it was recently proven that gauged PEPS are the most general gauge invariant states on the lattice we suspect that our states are the most general gauge invariant states in the continuum. A rigorous proof of this statement is left for future work.

Another possible future direction is to use these new states as a numerical ansatz for the analysis of gauge theories directly in the continuum. We explored this possibility by computing the generating functionals associated to these states in the scenario where they are perturbatively close to being Gaussian. Furthermore, we provided a  transfer matrix methodology to map the computation of the D-dimensional generating functional to the optimization of a D-1-dimensional eigenvalue problem. Since the 0 dimensional problem is trivial this allows, in principle, full computation of the generating functional. 

Crucially, we note that this generating functional may contain divergences that must be regularized. We discussed that this may be done perturbatively through a UV-cutoff, counterterms or through dimensional reduction the built in regulator of CMPS. Most notable this dimensional reduction does not break the gauge symmetry. 

Furthermore, recently it has also been understood \cite{Kull_Classification_of_matrix_product_states_with_a_local_gauge_symmetry, Blanik_Internal_structure_gauge_invariant_PEPS} that it is possible to exactly contract  pure-matter continuous tensor networks if the Hamiltonian of the transfer matrix is an integrable model. In future works we wish to investigate how this situation is changed with the inclusion of gauge degrees of freedom. 

Finally, the presented approach can also be used to take continuum limits of gauged matrix product operators. Since gauged matrix operators are related to dualities on the lattice, this opens a window towards studying dualities in the continuum. A first step in this direction has already been taken in \cite{Roose_T-duality_and_bosonization_as_examples_of_continuum_gauging_and_disentangling}. In the same spirit it will be interesting to study the fusion category of gauged continuous tensor networks.

{\color{red}}

\section*{Acknowledgements}
The authors acknowledge valuable discussions with Karan Tiwana, Albert Gasull Celades, Jose Garre Rubio, Jutho Haegeman,  Marco Rigobello and Beno\^it Tuybens. We would also like to thank the anonymous referee for their insightful comments and valuable suggestions that helped improve this manuscript This research is funded by the European Union (ERC, OverSign, 101122583). Views and opinions expressed are however those of the author(s) only and do not necessarily reflect those of the European Union or the European Research Council. Neither the European Union nor the granting authority can be held responsible for them. We acknowledge the support of the Israel Science Foundation (Grant No. 374/24).

\bibliography{biblio}

\end{document}